\documentclass[pra,amsmath,amssymb,twocolumn]{revtex4}

\input epsf.tex
\usepackage{graphicx}
\usepackage{amsthm}

\newcommand{\be}{\begin{equation}}
\newcommand{\ee}{\end{equation}}
\newcommand{\ba}{\begin{eqnarray}}
\newcommand{\ea}{\end{eqnarray}}
\newcommand{\ban}{\begin{eqnarray*}}
\newcommand{\ean}{\end{eqnarray*}}

\newcommand{\sandwich}[3]{\mbox{$ \langle #1 | #2 | #3 \rangle $}}
\newcommand{\ket}[1]{\mbox{$ | #1 \rangle $}}

\newcommand{\demi}{\frac{1}{2}}
\newcommand{\compl}{\begin{picture}(8,8)\put(0,0){C}\put(3,0.3){\line(0,1){7}}\end{picture}}

\newcommand{\one}{\leavevmode\hbox{\small1\normalsize\kern-.33em1}}

\begin{document}

\title{On the local and non-local content of bipartite qubit and qutrit correlations} \author{Valerio Scarani}
\affiliation{Centre for Quantum Technologies, National
  University of Singapore, Singapore} \date{\today}
\begin{abstract} The local and non-local contents of non-local probability distributions are studied using the approach of Elitzur, Popescu and Rohrlich [Phys. Lett. A \textbf{162}, 25 (1992)]. This work focuses on distributions that can be obtained by single-copy von Neumann measurements on bipartite quantum systems. For pure two-qubit states $\ket{\Psi(\theta)}=\cos\theta\ket{00}+\sin\theta\ket{11}$ (with $\cos\theta\geq\sin\theta$), the local content of the corresponding probability distribution is found to lie between $1-\sin2\theta$ and $\cos2\theta$. For the family $\ket{\Psi(\gamma)}= \big(\ket{00}+\ket{11}+\gamma \ket{22}\big)/\sqrt{2+\gamma^2}$ of two-qutrit states, non-zero local content is found for $\gamma>2$.
  
\end{abstract}
\maketitle

\section{Introduction} Since the 1930's, the entanglement between sub-systems of a composite quantum system has been recognized as one of the main features of the quantum description of nature \cite{epr,schro}. With the advent of quantum information and the interaction between physicists and computer scientists, entanglement has been recognized as a resource for communication and information processing. When local measurements are performed separately on each sub-system, correlations between the outcomes are observed, which in general cannot be ascribed to shared randomness \cite{bell}. This situation is now routinely referred to as \textit{quantum non-locality}. In addition to affecting our view of nature, this counter-intuitive feature can have an operational meaning in applied physics, as demonstrated recently in the context of quantum cryptography \cite{devindep}.

This paper builds on an idea Elitzur, Popescu and Rohrlich \cite{epr2}, referred to from now on as EPR2 for obvious reasons. EPR2 asked whether, in an experiment involving measurements of several photon pairs, one can consider that only some pairs behave non-locally, while the others would give rise to purely local correlations. This is a typical ``metaphysically-oriented'' formulation of the question: it proposes to explore the validity of an alternative description of nature. The same idea can be recast in the ``resource-oriented'' style of quantum information: one then asks if, in order to distribute some given non-local correlations, one can use non-local resources only partially and use shared randomness for the other instances. Yet another formulation of the same problem would consist in asking if a given non-local correlation is fully non-local or rather has a local content.

In Section \ref{secepr}, I review the EPR2 approach to non-locality and the few known results. In Sections \ref{secqubit} and \ref{secqutrit}, I present new results for the particular cases of the correlations arising out of single-copy von Neumann measurement on pure two-qubit, respectively two-qutrit, quantum states.

\section{The EPR2 approach to non-locality}\label{secepr}

\subsection{Local and non-local distributions}

For definiteness, in this paper I shall restrict to the bipartite case, i.e. the case where the global system consists of two sub-systems. In order to measure correlations, the two sub-systems are sent each to a different observer, called Alice and Bob. On her sub-systems, Alice can perform any measurement $A$ in a given set $\cal A$; the outcome of $A$ is labeled $r_A\in R_A$. Similarly, on his sub-system, Bob can perform any measurement $B$ in a given set $\cal B$; the outcome of $B$ is labeled $r_B\in R_B$. After repeating this procedure a large number of times, Alice and Bob together can estimate the family of bipartite probability distributions
\ba\left\{P(r_A,r_B|A,B)\big| r_A\in R_A, r_B\in R_B, A\in{\cal A},B\in{\cal B}\right\}\,.\ea
Such a family will be called \textit{local} if the correlations can be ascribed to shared randomness, i.e. if there exist a measure $\mu$ on an additional parameter $\lambda$ such that
\ba
P(r_A,r_B|A,B)&=&\int d\mu(\lambda)\,P(r_A|A,\lambda)P(r_B|B,\lambda)\label{plgen}
\ea
for all $A\in{\cal A}$ and $B\in{\cal B}$. Since locality is a condition on all possible measurements, its converse, namely \textit{non-locality}, can be demonstrated by checking only a subset of the possible measurements. The traditional approach to non-locality, generalizing the pioneering work of Bell \cite{bell}, is based on precisely that observation. Indeed, for any fixed number of measurements $(m_A,m_B)$ on each sub-system, there exist a set of constraints (Bell-type inequalities) that any local probability distribution (\ref{plgen}) must fulfill. One then checks if there exists subsets $\left\{A_i\right\}_{i\in I={1,...,m_A}}\subseteq {\cal A}$ and $\left\{B_j\right\}_{j\in J={1,...,m_B}}\subseteq {\cal B}$ such that the finite family $\left\{P_Q(r_A,r_B|A_i,B_j)\right\}_{i\in I,j\in J}$ violates these inequalities. If a violation happens, $P(r_A,r_B|A,B)$ is provably \textit{non-local}.

The fact that, on some states, non-locality can be checked with a finite number of measurements is crucial to demonstrate non-local correlations in an experiment. The EPR2 approach takes a different stand and aims at taking into account the whole set of possible measurements. Before discussing it, we shall remind the basic notations of quantum correlations.

\subsection{Quantum correlations}

In this paper I restrict quantum measurements to von Neumann measurements. Let ${\cal{H}}=\compl^{d_A}\otimes \compl^{d_B}$ be the Hilbert space that describes the system. The probability that the measurement of $A$ and $B$ performed on the state $\rho$ will lead to the outcomes $(r_A,r_B)$ is \ba
P_Q(r_A,r_B|\rho;A,B)&=&\mbox{Tr}\left({\Pi_{r_A}\otimes \Pi_{r_B}\,\rho }\right)\,,\label{eqpq}\ea where $\Pi_{r}$ is the projector on the subspace associated to the measurement result $r$. For any state $\rho$, the family of joint probability distributions
\ban
\Big\{P_Q(r_A,r_B|\rho;A,B)\Big| r_A\in R_A,r_B\in R_B, A\in{\cal A},B\in{\cal B}\Big\}
\ean can be either local or non-local. For simplicity, we shall say that \textit{the state $\rho$ is local} or \textit{non-local} in either case. We stress that, in this paper, this refers to the probability distribution obtained through \textit{single-copy von Neumann measurements}: some entangled states like Werner's states are local according to our definition \cite{werner} but can produce non-local correlations under generalized measurements \cite{popescu,horodecki}; by generalizing further the measurement procedure, it has been shown recently that any bipartite entangled state exhibits some non-local features \cite{mas08}.

For the goal of assessing the non-locality of the joint probability distribution, it is not restrictive to consider non-degenerate measurements and to assume that the sets of possible outcomes, i.e. the possible values of $r_A$ and $r_B$, are the same for all measurements \footnote{Indeed, one can always see a measurement as consisting of two steps: (i) projection on a basis; (ii) attribution of a numerical value for the outcome associated to each element of the basis. This numerical value is purely a label: any choice is valid, provided it does not erase information,as it would be the case if the same value would be associated to more than one outcome. For instance, one can choose $R_A=\{1,2,...,d_A\}$ and $R_B=\{1,2,...,d_B\}$.}. In particular, we can consider ${\cal A}$ (respectively, ${\cal B}$) to be the set of all possible orthonormal bases in $\compl^{d_A}$ (respectively, $\compl^{d_B}$). Obviously, there are infinitely many elements in these sets.

\subsection{EPR2 approach on bipartite quantum correlations}

The \textit{EPR2 approach} consists in considering the decompositions of $P_Q$ as the convex sum of a local ($P_L$) and a non-local ($P_{NL}$) probability distribution:
\ba
P_Q(r_A,r_B|\rho;A,B)&=& p_L(\rho)\,P_L(r_A,r_B|\rho;A,B)\nonumber\\
&+&[1-p_L(\rho)]\,P_{NL}(r_A,r_B|\rho;A,B)\label{epr2generic}
\ea
where the equality must hold for all outcomes $r_A,r_B$ and for all possible measurements $A,B$, and where the weight $p_L(\rho)\in[0,1]$ of the local component is required to be independent of the measurements and of the outcomes. The non-local component $P_{NL}$ is not restricted to come from the measurement of a quantum system \footnote{Note also that, if $\rho$ is pure, $P_L$ and $P_{NL}$ cannot be obtained from the measurement of a quantum state \textit{with the same dimensionality}, otherwise one would be able to decompose a pure state as a convex combination of other states.}. By construction, it satisfies the no-signaling condition, because $P_Q$ and $P_L$ do: indeed, the marginal distribution on Alice's side $P_{NL}(r_A|\rho;A,B)=[P_Q(r_A|\rho;A)-p_L(\rho)P_L(r_A|\rho;A)]/[1-p_L(\rho)]$ is independent of Bob's measurement $B$; and symmetrically for the marginal distribution on Bob's side.

The previous requirements do not single out a unique convex combination (\ref{epr2generic}): for instance, if $P_Q$ is non-local, the trivial choice $P_Q=0\times P_L+1\times P_{NL}$ is always possible. The EPR2 approach comes therefore with a natural \textit{optimization problem} associated to it: find the decomposition that maximizes the weight of the local component. We shall use the notation $p^{max}_L(\rho)$ for this optimal value, that can be interpreted as the \textit{local content} of $P_Q(r_A,r_B|\rho;A,B)$. By definition, the state $\rho$ is local (with respect to single-copy von Neumann measurements) if and only if $p^{max}_L(\rho)=1$.

Apart from defining the local content of a probability distribution, the EPR2 approach is naturally related to a question that has attracted a lot of interest recently, namely, the simulation of quantum correlations with other non-local resources \cite{brassard,toner,cerf, brunner,degorre}: indeed, assuming that shared randomness is a free resource, only $P_{NL}$ has to be simulated. I shall come back to this point in \ref{sspnl}.

\subsection{Known results on $p^{max}_L(\rho)$}

Very few exact results and even bounds are known on $p^{max}_L(\rho)$. In their seminal paper, EPR2 studied the family of two-qubit pure states $\cos\theta\ket{00}+\sin\theta\ket{11}$ and found an explicit decomposition that gives the lower bound
\ba
p^{max}_L(\theta)&\geq &\frac{1}{4}(1-\sin2\theta)\,.\label{epr2bound}
\ea
For the maximally entangled state, and under a reasonable continuity assumption, they could prove that the bound is tight: the singlet state of two qubits is fully non-local.

Recently, Barrett, Kent and Pironio found that an upper bound on $p^{max}_L(\rho)$ can be obtained from any Bell-type inequality \cite{bkp}. The argument goes as follows. Let $I$ be a linear combination of probabilities associated to different measurements that defines a Bell-type inequality, i.e. $I\leq I_L$ for all local probability distributions. Denote by $I(\rho)$ the quantum value on the state $\rho$ for the best choice of measurement settings, and $I_{NS}>I_L$ the maximal value that $I$ can take over all no-signaling distributions. Then $I(\rho)\leq p_L(\rho)I_L+[1-p_L(\rho)]I_{NS}$, i.e.
\ba
p^{max}_L(\rho)&\leq& \left[I_{NS}-I(\rho)\right]/\left[I_{NS}-I_L\right]\,.\label{boundineq}
\ea
In particular, if there exist an inequality for which $I(\rho)=I_{NS}$, then $p^{max}_L(\rho)=0$. With this argument, they proved conclusively that $p^{max}_L(\psi)= 0$ for the two-qubit singlet and extended the conclusion to maximally entangled states of arbitrary dimensionality.

To my knowledge, no other papers address the question of finding $p_L(\rho)$ with reference to the EPR2 paper \cite{epr2}. Obviously, the characterization of the states that produce fully local correlations can be seen as a particular instance of the EPR2 optimization problem, in which one asks whether $p^{max}_L(\rho)=1$ for a given state. This particular problem has received significant attention. For pure states, $p^{max}_L(\psi)=1$ holds if and only if the state is not entangled \cite{gisin,gisper,pr92}.However, there exist examples of mixed states that are entangled, but for which $p^{max}_L(\rho)=1$ holds \cite{werner,terhal,acin}, in some case also under single-copy POVMs \cite{barrett}. The general problem is unsolved, the non-locality of bipartite bound-entangled states being a specially intriguing open issue.

In summary, the information available in the literature is:
\begin{itemize}
\item $p^{max}_L(\psi)=0$ holds for all bipartite \footnote{In the multipartite case, $p^{max}_L(\rho)=0$ holds for all states that exhibit a Greenberger-Horne-Zeilinger type of non-locality. There are examples of such states that are not maximally entangled and actually mixed [V. Scarani, A. Ac\'{\i}n, E. Schenck, M. Aspelmeyer, Phys. Rev. A \textbf{71}, 042325 (2005)]. Note that, for a $N$-partite state, $p^{max}_L(\rho)=0$ does not necessarily imply genuine $N$-party non-locality.} maximally entangled states;
\item $p^{max}_L(\rho)=1$ holds for separable states but also for some mixed entangled states.
\item For intermediate cases, we have an explicit lower bound for pure states of two qubits (\ref{epr2bound}) and an upper bound derivable from Bell-type inequalities (\ref{boundineq}).
\end{itemize}

In this paper, I present an improved lower bound and a new upper bound for the local content $p^{max}_L(\theta)$ on the family of pure two-qubit states (Section \ref{secqubit}) and the first example of a lower bound on the local content of pure two-qutrit states (Section \ref{secqutrit}).

\section{Pure two-qubit states}\label{secqubit}

In this Section, we study the non-locality of pure non-maximally entangled states of two qubits
\ba
\ket{\Psi(\theta)}&=& \cos\theta\ket{00}+\sin\theta\ket{11}\label{famqubit}
\ea
where we can choose $\cos\theta\geq \sin\theta$ without loss of generality. Although this is the simplest possible family of states, its non-locality is not completely understood: for instance, it is known \cite{brunner} that weakly entangled states cannot be simulated by a single use of the non-local resource called Popescu-Rohrlich (PR) box \cite{PR}, whereas the maximally entangled state can \cite{cerf}.

A local von Neumann measurement is labeled by a unit vector $\vec{v}$; the projector corresponding to the outcome $r=\pm 1$ is $\Pi_r=\demi\left(\one+\,r\,\vec{v}\cdot\vec{\sigma}\right)$ with $\vec{\sigma}$ the vector of Pauli matrices. For each $\ket{\Psi(\theta)}$, the family of quantum probability distribution under study is therefore
\ba
P_Q&=&\frac{1}{4}\big[1+r_Aca_z + r_Bcb_z+ r_Ar_B E_Q(\vec{a},\vec{b})\big]\label{pq}
\ea
with $E_Q(\vec{a},\vec{b})=a_zb_z+s(a_xb_x-a_yb_y)$ and where we have written $\cos2\theta=c$ and $\sin2\theta=s$.

\subsection{Improved lower bound}

\subsubsection{Decomposition and bound on $p^{max}_L(\theta)$}

EPR2 assumed $P_L=\frac{1}{4}\big[1+r_A\mbox{sign}(a_z)\big]\big[1+r_B\mbox{sign}(b_z)\big]$, which, inserted in (\ref{epr2generic}), defines a valid $P_{NL}$ as long as $p_L\leq\frac{1}{4}(1-\sin2\theta)$, leading to the bound (\ref{epr2bound}). This shows that the correlations arising from single-copy measurements on non-maximally entangled state of two qubits are not fully non-local. However, the decomposition is obviously not optimal: in particular, in the limit $\theta\rightarrow 0$ of the product state, the bound on $p^{max}_L(\theta)$ tends to $\frac{1}{4}$, while we know that $p^{max}_L(\theta=0)=1$ must hold because a product state is local.

A significant improvement can be obtained while still keeping the product form for the local probability distribution
\ba P_L &=& \frac{1}{4}\left[1+r_A f(a_z)\right]\,\left[1+r_B
f(b_z)\right]\,:\label{plnew}\ea indeed, with the choice $f(x)=\mbox{sign}(x)\,\min\big(1,\frac{c}{1-s}|x|\big)$, this leads to a valid $P_{NL}$ provided $p_L\leq 1-\sin2\theta$, leading to the improved lower bound\ba
p^{max}_L(\theta)&\geq &1-\sin2\theta\,.\label{pLbest}
\ea We note that this is the largest possible value if $P_L$ is supposed to depend only on $a_z$ and $b_z$ \footnote{The elegant proof that follows is due to Cyril Branciard and reproduced with his permission.}. Suppose indeed that $P_L(r_A,r_B|\vec{a},\vec{b})=P_L(r_A,r_B|a_z,b_z)$. Then let's take $\vec{a}=r_A\hat{x}=(r_A,0,0)$ and $\vec{b}=-r_B\hat{x}=(-r_B,0,0)$ where $\hat{x}$ is the unit vector in the $x$ direction. The quantum prediction is $P_Q(r_A,r_B|r_A\hat{x},-r_B\hat{x})=\frac{1}{4}(1-s)$, while by assumption $P_L(r_A,r_B|r_A\hat{x},-r_B\hat{x})=P_L(r_A,r_B|0,0)$. Using (\ref{epr2generic}), we obtain $(1-p_L)P_{NL}(r_A,r_B|r_A\hat{x},-r_B\vec{x})=\frac{1}{4}(1-s)-p_LP_L(r_A,r_B|0,0)$, and, for $P_{NL}$ to be correctly defined, the right-hand side expression must be non-negative for all $(r_A,r_B)$. By summing over $(r_A,r_B)$ we obtain $1-s-p_L\geq 0$.

\subsubsection{Study of $P_{NL}$}\label{sspnl}

In order to obtain the lower bound on $p^{max}_L(\theta)$, the only requirement is that $P_{NL}$ is a valid probability distribution. However, as mentioned above, the definition of the local content of a probability distribution is not the only point of interest in the EPR2 approach to non-locality. In particular, a detailed study of $P_{NL}$ may help in the task of simulating quantum correlations. For the decomposition that we are discussing, the non-local component reads \ba
P_{NL}&=&
\frac{1}{4}\left[1+r_A F(a_z) + r_B F(b_z)+r_Ar_B
G(\vec{a},\vec{b})\right]\label{pnl}\ea with $F(x)=\frac{1}{s}[cx-(1-s)f(x)]$ and $G(\vec{a},\vec{b})=a_xb_x\,-\,a_yb_y\,+\, \frac{1}{s}[a_zb_z-(1-s)f(a_z)f(b_z)]$. Here are some features of this family of $P_{NL}$:
\begin{itemize}  
\item No $P_{NL}$ in this family violates the Clauser-Horne-Shimony-Holt inequality \cite{chsh} beyond the quantum bound \cite{tsirelson}.
\item If both Alice's and Bob's settings lie in a band across the equator of the Bloch sphere defined by $|a_z|,|b_z|\leq \frac{1-s}{c}$, $P_{NL}$ takes the simple form $\frac{1}{4}\big[1+r_Ar_B (a_xb_x-a_yb_y-a_zb_z)\big]= \frac{1}{4}\big[1+r_Ar_B \vec{a}\cdot\tilde{b}\big]$ with $\tilde{b}=(b_x,-b_y,-b_z)$: the local distributions are random and the correlations are like a scalar product. Thus, under the promise that the measurement settings are close enough to the equator \footnote{Note that the band defined by $|a_z|,|b_z|\leq \frac{1-s}{c}$ becomes \textit{narrower} when approaching the maximally entangled state. From this perspective, the decomposition we have just found behaves in a discontinuous way: (i) for the maximally entangled state, $P_{NL}=P_Q$ has random marginals and correlations defined by a scalar product; (ii) for highly but non-maximally entangled states ($\theta=\frac{\pi}{4}-\varepsilon$), such a simple form is however retained only for a narrow band of settings; the band increases when going to weakly entangled states, filling ultimately again the whole sphere for $\theta=0$, in which case however the weight of $P_{NL}$ is zero.}, $P_{NL}$ can be simulated by a single use of a PR box adapting the protocol in Ref.~\cite{cerf}. In other words, measurement settings that are close to the poles of the Bloch sphere must be used to verify that a single PR box is not enough to simulate the correlations of non-maximally entangled states \cite{brunner}.
\end{itemize}
Based on (\ref{pnl}), it has been possible to derive the first simulation of the correlations of pure two-qubit states that uses only no-signaling resources and such that the amount of non-local resources decreases with the degree of entanglement \cite{bgps08}. This is the first achievement of the EPR2 approach to non-locality in the field of simulation of quantum correlations.

\subsection{Upper bound based on the chained inequality}

An upper bound on $p^{max}_L(\theta)$ is provided by (\ref{boundineq}) under the choice of a Bell-type inequality. Following \cite{bkp}, we consider the family of ``chained inequalities'' $I^{(N)}$ with $N$ settings for both Alice and Bob \cite{pearle} and find $I^{(N)}(\theta)$, the maximal violation that can be obtained for the state $\ket{\Psi(\theta)}$. This involves an optimization on the settings, the last step of which has been done numerically 
(see Appendix \ref{appa} for the details). The resulting bound is $p^{max}_L(\theta)\leq \cos2\theta(1+\delta^{(N)})$ where $\delta^{(N)}$ is found to be decreasing with $N$ and $\delta^{(40)}\approx 10^{-5}$. In a compact form therefore,
\ba
p^{max}_L(\theta)&\lesssim& \cos2\theta\,.\label{conjpl}
\ea

\subsection{Pure two-qubit states: summary}

\begin{figure}
\includegraphics[scale=0.55]{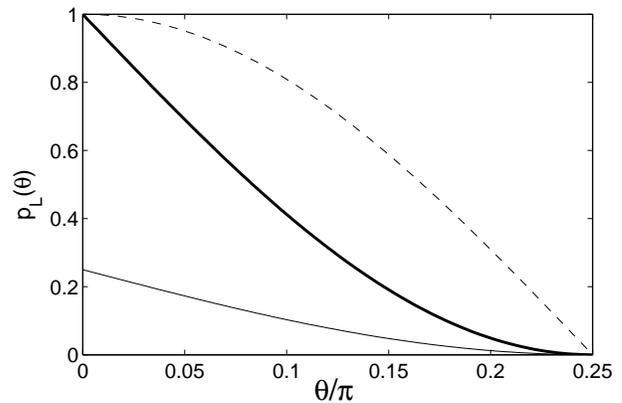}
\caption{Bounds on $p^{max}_L(\theta)$ as a function of $\theta$ for pure two-qubit states (\ref{famqubit}) --- in particular, $\theta=0$ is the separable state, $\theta=\frac{\pi}{4}$ is the maximally entangled state. Thin full line: the EPR2 lower bound $\frac{1}{4}(1-\sin2\theta)$. Thick full line: the lower bound (\ref{pLbest}) obtained in this paper. Dashed line: the numerical upper bound (\ref{conjpl}).} \label{figqubit}
\end{figure}

Figure \ref{figqubit} summarizes the present knowledge on the local content of the probability distribution associated to single-copy von Neumann measurements of pure two-qubit states. On the one hand, the lower bound on $p^{max}_L(\theta)$ has been significantly improved with respect to the original EPR2 paper \cite{epr2} using a decomposition that has later proved useful in the task of simulating entanglement \cite{bgps08}. On the other hand, an upper bound has been derived for all $\theta$ using the techniques of \cite{bkp}. The lower and the upper bound coincide only for the maximally entangled state and for the product state: the exact value of $p^{max}_L(\theta)$ is still unknown for all non-maximally states.

\section{Pure two-qutrit states}
\label{secqutrit}

I turn now to the first study of the local content of pure state of two qutrits. I restrict to a one-parameter family of states
\ba
\ket{\Psi(\gamma)}&=& \frac{1}{\sqrt{2+\gamma^2}}\big(\ket{00}+\ket{11}+\gamma \ket{22}\big)\label{psigamma}
\ea
that appears as relevant in several studies of non-locality \cite{adgl02,agg05,scarani}. It is already known that $p^{max}_L(\gamma=0)=p^{max}_L(\gamma=1)=0$, because these are maximally entangled states \cite{bkp}. I am going to construct an explicit form of $P_L$, which allows to show the lower bound
\ba
p^{max}_L(\gamma)>0 &\mathrm{for} &\gamma\gtrsim 2\,.\label{boundqutrit}
\ea

The construction is the generalization of the EPR2 one for qubits. There, since the state (\ref{famqubit}) is biased towards $\ket{00}$, the most probable local outcomes are $r_A=\mbox{sign}(a_z)$ and $r_B=\mbox{sign}(b_z)$. EPR2 chose $P_L$ to be the distribution that produces those outcomes deterministically. The same intuition can be applied here if $\gamma>1$, because the state is then biased towards $\ket{22}$. Explicitly, since a generic local measurement is described by a unitary operation $U_A\otimes U_B$ followed by measurement in the computational basis, $P_L$ is constructed as follows:
\begin{enumerate}
\item On Alice's side, compute $p_i=|\sandwich{i}{U_A}{2}|^2$ for $i=0,1,2$; denote by $i(A)$ the value of $i$ that gives the largest probability, i.e. $p_{i(A)}=\max_ip_i$.
\item Similarly, on Bob's side, compute $p_j=|\sandwich{j}{U_B}{2}|^2$ for $j=0,1,2$; denote by $j(B)$ the value of $j$ that gives the largest probability, i.e. $p_{j(B)}=\max_jp_j$.
\item Then
\ba
P_L(r_A,r_B|\gamma,A,B)&=&\delta_{r_A,i(A)}\,\delta_{r_B,j(B)}\,.
\ea Note that $P_L$ is independent of $\gamma$
\end{enumerate}

\begin{figure}
\includegraphics[scale=0.55]{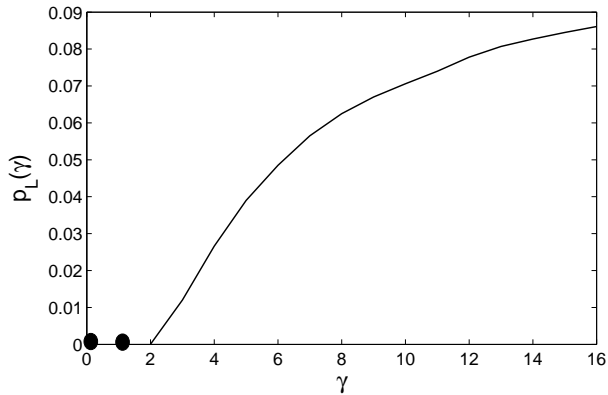}
\caption{Present knowledge about $p^{max}_L(\gamma)$ for the family (\ref{psigamma}) of pure two-qutrit states --- in particular, $\gamma=0$ is the maximally entangled state of two qubits, $\gamma=1$ is the maximally entangled state of two qutrits, and the state becomes separable in the limit $\gamma\rightarrow\infty$. The black dots represent the proved result $p^{max}_L(\gamma=0)=p^{max}_L(\gamma=1)=0$. The line is a lower bound on $p^{max}_L(\gamma)$ obtained for the choice of $P_L$ described in the text (numerical study, the precision on the bound is approximately $10^{-5}$).} \label{figtrit}
\end{figure}

Given this $P_L$ and $P_Q$, one computes the largest weight of the local component such that $P_{NL}$ is still a valid probability distribution. This study was done using numerical minimization algorithms in Matlab. The result is plotted in Fig.~\ref{figtrit}. We see in particular that the lower bound (\ref{boundqutrit}) holds as announced.

Among the several open questions, I single out the study of the region $0<\gamma<1$. This region contains the state that gives the maximal violation of the unique Bell inequality with two settings per site ($\gamma=\frac{\sqrt{11}-\sqrt{3}}{3}\approx 0.792$) \cite{adgl02} and other meaningful states \cite{agg05,scarani}.
It is therefore a promising region in which to investigate if pure \textit{non-maximally} entangled states can generate a fully non-local probability distribution \footnote{For the whole region $0<\gamma<1$, the upper bound (\ref{boundineq}) obtained using the chained inequality gives a strictly positive number (S. Pironio, private communication).}.

\section{Conclusion}

The local and non-local content of a probability distribution arising from single-copy von Neumann measurements on bipartite quantum states have been defined after the ideas of Elitzur, Popescu and Rohrlich \cite{epr2}. Contrary to Bell's inequalities, that check non-locality on a subset of the measurement settings \footnote{It is an interesting open question, whether $p^{max}_L(\rho)<1$ implies the possibility of checking non-locality with a \textit{finite} number of settings.}, this approach considers the whole set of possible measurements. As such, it is related to the issue of simulating quantum correlations with non-local resources.

I have presented improvements on the original study for pure states of two qubits, as well as the first application of the approach for pure non-maximally entangled states of two qutrits.

\section*{Acknowledgements} I acknowledge discussions with A. Ac\'{\i}n, C. Branciard, N. Brunner, N. Gisin, Y.-C. Liang, S. Pironio and S. Popescu. This work is supported by the National Research Foundation and Ministry of Education, Singapore.

\begin{appendix}

\section{On the optimization of the chained inequality with von Neumann measurements on pure two-qubit states}\label{appa}

This appendix is devoted to the chained inequality \cite{pearle,bkp} and its optimization with von Neumann measurements on pure two-qubit states. The inequality with $N$ settings for Alice and as many for Bob can be written as a correlation inequality:
\ba
I^{(N)}&=&A_1B_1+B_1A_2+A_2B_2+\nonumber\\&&...+A_NB_N-B_NA_1\,\leq\,2(N-1).
\ea The no-signaling bound is $I^{(N)}_{NS}=2N$ and can be reached with a single PR-box by the following prescription: Alice inputs $x=1$ for $A_1$ and $x=0$ otherwise, Bob inputs $y=1$ for $B_N$ and $y=0$ otherwise. The corresponding Bell operator associated to von Neumann measurements on qubits is, with the notation $\vec{a}\otimes \vec{b}\equiv \big(\vec{a}\cdot\vec{\sigma}\big)\otimes \big(\vec{b}\cdot\vec{\sigma}\big)$:
\ba
{\cal I}^{(N)}&=&\sum_{k=1}^{N-1}\left(\vec{a}_k+\vec{a}_{k+1}\right)\otimes \vec{b}_k+ \left(\vec{a}_N-\vec{a}_{1}\right)\otimes \vec{b}_N\nonumber\\
&=& \sum_{k=1}^{N-1}A_k \vec{c}_k\otimes \vec{b}_k+ A_N\vec{c}_N\otimes \vec{b}_N
\ea where $A_k \vec{c}_k=\vec{a}_k+\vec{a}_{k+1}$ and $A_N \vec{c}_N=\vec{a}_N-\vec{a}_{1}$.

The goal is to find the measurement directions that maximize 
\ba I^{(N)}(\theta)&=&\sandwich{\Psi(\theta)}{{\cal I}^{(N)}}{\Psi(\theta)}\ea for $\ket{\Psi(\theta)}$ given in (\ref{famqubit}). Because $\sandwich{\Psi(\theta)}{\vec{a}\otimes \vec{b}}{\Psi(\theta)}=a_zb_z+\sin2\theta\big(a_xb_x-a_yb_y\big)$, the optimal choice of settings is such that
\ba
\vec{a}_k=\cos\alpha_k\hat{z}+\sin\alpha_k\hat{x}&\mathrm{and} & \vec{b}_k=\vec{c}_k
\ea for all $k$. So we are left with $N$ parameters to optimize, the angles $\alpha_k$ in the $(x,z)$ plane.

\begin{figure}
\includegraphics[scale=0.55]{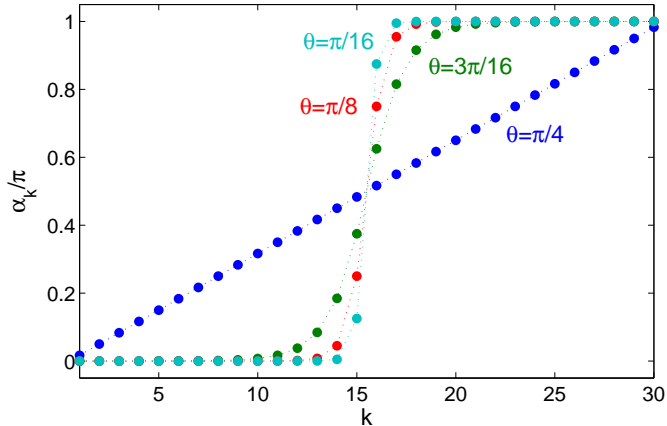}
\caption{(Color online). Optimal measurements settings $\alpha_k$ ($k=1,...,30$) for the chained inequality $I^{(30)}$ and different values of $\theta$.} \label{figsett}
\end{figure}

For the maximally entangled state, the solution is known analytically: $I^{(N)}(\frac{\pi}{4})=2N\cos\left(\pi/2N\right)$ for $\alpha_k=\frac{k-1}{N}\pi$. This choice of settings is however provably sub-optimal for non-maximally entangled states. Resorting to numerical methods, one finds that several sets of angles provide the same value of $I^{(N)}(\theta)$ within numerical precision; in particular, one can find a set with an elegant symmetry of reflection around the $x$-axis, i.e. with $\alpha_{N-k+1}=\pi-\alpha_k$. The result of the numerical optimization under this symmetry is shown in Fig.~\ref{figsett}. For $\theta=\frac{\pi}{4}$, the settings are equally spaced as already discussed; for weakly entangled states, the settings tend to concentrate close to $\alpha=\pm\hat{z}$, with a sharp transition between these two values | not an unexpected behavior, since the state has a large $\ket{00}$ component.

\end{appendix}

\end{document}